\documentclass[12pt]{article}

\newcommand{\pder}[2]{{\partial #1 \over \partial #2}}

\def\gsim{\;\raise0.3ex\hbox{$>$\kern-0.75em\raise-1.1ex\hbox{$\sim$}}\;}
\def\lsim{\;\raise0.3ex\hbox{$<$\kern-0.75em\raise-1.1ex\hbox{$\sim$}}\;}
\def\vector#1{{\bf #1}}

\def\pref#1{(\ref{#1})}
\def\ignore#1{}  
\def\roughly#1{\mathrel{\raise.3ex\hbox{$#1$%
\kern-.75em\lower1ex\hbox{$\sim$}}}}
\def\lsim{\roughly<}
\def\gsim{\roughly>}

\begin{document}

\centerline{\Large\bf Neutrino kinetics in a magnetized dense plasma}

\begin{center}
{\large\bf V.N. Oraevsky, V.B. Semikoz}
\vskip0.5cm
              The Institute of Terrestrial Magnetism, Ionosphere and \\
              Radio Wave Propagation of the Russian Academy of Sciences,\\
               IZMIRAN, Troitsk, Moscow region, 142190, Russia
\end{center}

\abstract{The relativistic kinetic equations (RKE) for lepton plasma
in the presence of a strong external magnetic field are
derived in Vlasov approximation. The new RKE for the electron spin
distribution function includes the weak interaction with neutrinos
originated by the axial vector current ($\sim c_A$) and provided by the
parity nonconservation. In a polarized electron gas Bloch equation
describing the evolution of the magnetization density perturbation is
derived from the electron spin RKE being modified in the presence of
neutrino fluxes. Such modified hydrodynamical equation allows to obtain
the new dispersion equation in a magnetized plasma from which {\it the
neutrino driven instability of spin waves} can be found. It is shown that
this instability is more efficient e.g. in a magnetized supernova than
the analogous one for Langmuir waves enhanced in an isotropic plasma.  }


\newpage

\section{Introduction} When he conjectured the existence of the neutrino,
Pauli imposed very stringent bounds on its electrical neutrality.

Nevertheless, the direct (=weak) interaction of neutrino with
electrons and positrons shows a principle possibility of the interaction
of neutrino with the electromagnetic field. In vacuum, however, this
interaction is negligible. The situation is extremely changed in media with
free carriers of the electric charge such as a dense plasma of metals,
stars, the lepton plasma of the early Universe, etc.

It was shown in \cite{Oraevsky,Oraevsky2,Nieves} that the electrodynamics
of neutrinos is changed in such media comparing with the electrodynamics
of neutrinos  in vacuum  in such a way that the
contribution of the neutrino electromagnetic vertex in medium cannot be
regarded as a small correction like the case of radiative corrections in
vacuum. The main distinction is the appearance of {\it
the induced electric charge of neutrino} which is proportional to the
density of free carriers of electricity \footnote{There appears also {\it
induced magnetic moment of neutrino} \cite{Semikoz3}, cross-sections of
the neutrino scattering off nuclei are modified in plasmas especially in
low energy region \cite{Leinson} and so on.}. Namely this leads to the
new qualitative effect: the appearance of the long-distance forces
inevitably leads to the collective interactions of neutrino with an
ensemble of charged particles via electromagnetic field.

The approximation of the neutrino propagation in an isotropic plasma
considered in \cite{Oraevsky, Oraevsky2,Nieves,Semikoz} and then in
\cite{Bingham} has some natural bounds for applications in astrophysics.
Therefore one needs to have the self-consistent system of kinetic
equations of the most general kind applicable for other astrophysical
plasmas including magnetized stars.

Really this turns out to be a new problem unsolved before in the neutrino
kinetics while in the one-particle Schr\"{o}dinger equation approach
one can show the importance of the axial vector potential
for the neutrino propagation in a magnetized plasma ($V_A\sim
<\bar{\psi}_e\gamma_i\gamma_5\psi_e>\sim B_i$) that provides the
interaction of neutrino with the magnetic field {\bf B} without
any neutrino magnetic moment \cite{SemikozValle} and could explain, e.g.
the neutron star kick \cite{Kusenko}.

In the present work we take into account the spin
interactions of electrons that are important for neutrino propagation in a
dense degenerate electron gas polarized by the external magnetic field.

In subsection 2.1. we discuss the equilibrium state of an electron
gas polarized by the external magnetic field $\vector{B}_0\neq 0$.

Then in subsection 2.2 we present the full set of the coupled
Relativistic Kinetic Equations (RKE) for electrons and neutrinos in the
lepton plasma with an external magnetic field including the
self-consistent electromagnetic field contribution. These RKE's are derived
using the Bogolyubov method analogously to the way described in details
for an isotropic plasma in \cite{Semikoz}.

In subsection 2.3 we complete derivation of RKE's recasting
them for the gauge invariant distribution functions and prove the lepton
current conservation,\\ $\partial j^{(a)}_{\mu}(\vector{x}, t)/\partial
x_{\mu}=0$, in a magnetized plasma both for neutrinos and electrons.

We conclude this section deriving in subsection 2.4 the dispersion
equation for the lepton density perturbations in an isotropic plasma that
coincides with results \cite{Bingham,Bento}.

In section 3 we generalize the Bloch equation for the
evolution of the magnetization density perturbation in the presence of
neutrino fluxes. In section 4, neglecting spatial dispersion we
obtain the increment of spin waves excited by the neutrino beam and compare
it with the case of the neutrino driven instability of plasma waves in an
isotropic plasma.

In section 5 we discuss results and give some estimates of
relevant parameters in a polarized medium.

 \vskip 0.3cm \section{Relativistic Kinetic Equations for lepton plasma in
 Vlasov approximation}
The kinetic equations in the Standard Model (SM)
 are derived from the quantum Liouville equation for the non-equilibrium
statistical operator $\hat{\rho }(t)$ using the Bogolyubov method with the
total interaction Hamiltonian given by the Feynman diagrams for the
neutrino scattering off electrons and the usual ee-scattering in QED
\cite{Semikoz}. For a magnetized plasma we do not consider $\nu
\nu$-scattering and neglect also weak $ee$-scattering comparing with the
electromagnetic interaction of charged particles through the photon
exchange.

The one-particle density matrix
$\hat{f}^{(a)}_{r'r}(\vector{p}, \vector{x},t) =
\sum_{\vector{k}}e^{i\vector{k}\vector{x}}
Tr(\hat{\rho}(t)\hat{b}^{+(a)}_{\vector{p}-\vector{k}/2,r}
\hat{b}^{(a)}_{\vector{p}+\vector{k}/2, r'})$ is determined as the
statistical average of the
product of creation ($\hat{b}^{+(a)}_{\vector{p}r}$) and annihilation
($\hat{b}^{(a)}_{\vector{p'}r'}$) operators for the lepton $a=e, \nu$. In
the case of electrons it takes the form
$$\hat{f}^{(e)}_{\lambda'\lambda}(\vector{p}_e,\vector{x},t)=
f^{(e)}(\vector{p}_e,\vector{x},t)\frac{\delta_{\lambda'\lambda}}{2} +
S^{(e)}_i(\vector{p}_e,\vector{x},t)
\frac{(\sigma_i)_{\lambda'\lambda}}{2}~,
$$ where $\lambda=\pm 1$ is the
eigenvalue of the spin projection on an
external magnetic field. E.g. for the magnetic
field $\vector{B}= (0,0,B_0)$ only the spin component $\Sigma_z$
commutes with the Hamiltonian for electrons,
$[\Sigma_z,\hat{\cal{H}}^{(e)}]_=0$, hence $(\sigma_z)_{\lambda'\lambda}=
\lambda \delta_{\lambda'\lambda}$.
In the absence of neutrino magnetic moment the one-particle density
matrix for massless Dirac neutrinos  corresponds to the active
(left-handed, $r= -1$) neutrino distribution function only,
$$\hat{f}^{(\nu)}_{r'r} (\vector{q},\vector{x}, t)= \frac{(1 -
r)\delta_{r'r}}{2}f^{(\nu)} (\vector{q},\vector{x}, t)~,$$ where in an
uniform medium the helicity $r=\pm 1$ conserves,
$(\sigma_iq_i)_{r'r}=qr\delta_{r'r}$.

The number density distribution functions are related with the
lepton densities $n^{e}(\vector{x},t) = \int
(d^3p/(2\pi)^3)f^{(e)}(\vector{p}, \vector{x},t)$ for electrons and
$n^{\nu}(\vector{x},t) = \int (d^3q/(2\pi)^3) f^{(e)}(\vector{q},
\vector{x},t) $ for neutrinos.

\vskip 0.3cm \subsection{Equilibrium state in a polarized
electron gas}

In the linear approximation for a slightly inhomogeneous medium the
 distribution functions have the form \begin{eqnarray} \label{linear}
 &&f^{(\nu)}(\vector{q}, \vector{x},t) = f^{(\nu)}_0(\vector{q}) + \delta
f^{(\nu)}(\vector{q}, \vector{x},t)~,\nonumber\\&& f^{(e)}(\vector{p},
\vector{x},t) = f^{(e)}_0(\varepsilon_p) + \delta f^{(e)}(\vector{p},
\vector{x},t)~,\nonumber\\&& \vector{S}^{(e)}(\vector{p}, \vector{x},t) =
\vector{S}^{(e)}_0(\varepsilon_p) + \delta
 \vector{S}^{(e)}(\vector{p},\vector{x},t)~,  \end{eqnarray}
 where the neutrino background is given by $f^{(\nu)}_0(\vector{q})$
 not being in equilibrium with plasma environment (e.g. for the
powerful neutrino flux outside of the SN neutrinosphere), and we
 consider the uniform polarized equilibrium electron gas for which the
 one-particle Wigner density matrix takes the form \begin{equation}
 \label{background} \hat{f}^{(0e)}_{\lambda'\lambda}(\varepsilon_p) =
 \frac{\delta_{\lambda'\lambda}}{2}f^{(0)}(\varepsilon_p) +
 \frac{(\sigma_j)_{\lambda'\lambda}}{2}S_j^{(0)}(\varepsilon_p)~.
 \end{equation}
 Here  $f^{(0)}(\varepsilon_p)= g_e[\exp
 (\varepsilon_p - \zeta)/T) + 1]^{-1} $ is the equilibrium Fermi function;
$g_e = 2$ is the Lande factor;
\begin{equation}
 \label{polar}
 S_j^{(0)}(\varepsilon_p)= -
 \frac{\mid\mu_B\mid B^{(0)}_j }{\gamma}\frac{df^{(0)}(\varepsilon_p)}
 {d\varepsilon_p }
 \end{equation}
 is the equilibrium spin distribution; $\gamma= \varepsilon_p/m_e$ is
the electron gamma factor, and we use hereafter units
$\hbar=c=1$.

Note that in \pref{polar} we assumed the
quasi-classical electron spectrum in a realistic external magnetic field
$B_0$ obeying the inequality $eB_0\ll T^2$, when the Landau spectrum
\begin{equation}\label{Landau}\varepsilon (\lambda,p_z, n) = (m_e^2 +
 p_z^2 + \mid e\mid B_0(2n + 1 - \lambda))^{1/2}\end{equation}  reduces to  \begin{equation}
\label{spectrum1} \varepsilon_{p}(\lambda)= \varepsilon_p - \frac{\lambda
\mid e\mid B_0}{2\varepsilon_p}~, \end{equation} where $\lambda = \pm
1$ and we changed $\mid e\mid B_0(2n + 1)=p_{\perp}^2$ for large
Landau numbers $n\gg 1$. Thus, the full spectrum \pref{spectrum1}
contains the small paramagnetic (spin) correction to the continuous
spectrum $\varepsilon_p = \sqrt{p^2 + m_e^2}~.$

 One can easily check that in this quasi-classical limit
the exact expression for magnetization of the
electron gas, \begin{eqnarray} \label{exact} M_j^{(0)}=&&
 \mid \mu_B\mid<\bar{\psi}_e\gamma_j\gamma_5\psi>_0 =
\mid
\mu_B\mid\sum_{n=0}^{\infty}\frac{\mid
e\mid B_0}{(2\pi)^2}\int_{-\infty}^{+\infty}dp_zTr
[\sigma_j\hat{f}^{(0e)}(\varepsilon
(\lambda,p_z,n))]=\nonumber\\&& =\mid
\mu_B\mid\sum_{n=0}^{\infty}\frac{\mid
e\mid B_0\delta_{jz}}{(2\pi)^2}\int_{-\infty}^{+\infty}dp_z\sum_{\lambda}
\lambda {f}^{(0e)}_{\lambda \lambda}(\varepsilon(\lambda, p_z,n))~,
 \end{eqnarray} takes the
 quasi-classical form \begin{equation}\label{magnetization} M_j^{(0)}=
 \mid\mu_B\mid\int \frac{d^3p}{(2\pi)^3} S_j^{(0)}(\varepsilon_p) = -
 2\mu_B^2B_{0j}\int D(\varepsilon_p)
 \frac{df^{(0)}(\varepsilon_p)}{d\varepsilon_p} d\varepsilon_p~,
 \end{equation} where the spin distribution $S_j^{(0)}(\varepsilon_p)$ is
 given by eq.  \pref{polar} and $D(\varepsilon_p)= pm_e/(2\pi)^2$.  In
 particular, in the non-relativistic (NR) limit this background
magnetization corresponds to the spin paramagnetism of the free electron
gas in metal \cite{Kubo}.

 Note that for the degenerate electron gas   from
the quasi-classical Eq.  \pref{magnetization} one obtains the same value
 $M^{(0)}_z=\mid \mu_B\mid \mid e\mid B_0p_{F_e}/2\pi^2$ as it follows
 from the exact quantum eq.  \pref{exact} for an arbitrary strong magnetic
 field for which electrons populating the main Landau level (n=0)
contribute only.

Note also that this magnetization determines the axial vector
 potential of a probing neutrino in a magnetized plasma too
\cite{SemikozValle}, $$V_A= -G_F\sqrt{2}c_A\frac{q_j}{q}\frac{M_j^{(0)}}{
\mid \mu_B\mid}~,$$
that changes the spectrum of the ultrarelativistic neutrino,
$\varepsilon_q= q + V + V_A$, comparing with the
standard one in an isotropic medium (e.g. in the Sun), $\varepsilon_q=
q + V$, and modifies the neutrino oscillations in a magnetized SN described
by the one-particle Schr\"{o}dinger equation for two neutrino species
\cite{SemikozValle,Kusenko}, \begin{equation} i\left( \begin{array}{l}
\dot{\nu}_a\\ \dot{\nu}_b \end{array} \right) = \left( \begin{array}{cc} V
+ V_A -c_2\delta~~~ & s_2\delta \\ s_2\delta &0~ \end{array} \right)
\left( \begin{array}{c} \nu_a\\ \nu_b \end{array} \right)~, \label{master}
\end{equation}
where $c_2 = \cos 2\theta$, $s_2 = \sin 2\theta$, $\delta = \Delta
m^2/4E$ are the neutrino mixing parameters; $\nu_b$ is the sterile
neutrino wave function.

In our kinetic approach for the SM lepton plasma we do not consider
oscillations ($s_2=0$) assuming the massless spectrum $\varepsilon_q= q$
for active neutrino of the given kind $\nu_a$, $a = e, \mu, \tau$.

\subsection{The master kinetic equations for perturbations in plasma}
Within the linear approximation \pref{linear} the RKE for the Lorentz-invariant
Wigner number
density distribution functions $\delta f^{(a)}(\vector{p},\vector{x}, t) =
Tr\Bigl [\hat{\delta f}^{(a)}(\vector{p},\vector{x}, t ) \Bigr ] $
 take the covariant forms \begin{eqnarray}
\label{neutrino} &&q_{\mu}\pder{\delta f^{(\nu)}(\vector{q},\vector{x}, t
)}{x_{\mu}} + G_F\sqrt{2} c_V\int
\frac{d^3p_e}{(2\pi)^3}\frac{(p_e^{\mu}q_{\mu})}{\varepsilon_{p_e}}\pder
{\delta f^{(e)}(\vector{p}_e,\vector{x}, t  )}{x_j}
\pder{f^{(\nu)}_0(\vector{q})}{q_j} +\nonumber\\
&&+\frac{G_Fc_Am_eq^{\mu}}{\sqrt{2}}\int
\frac{d^3p_e}{(2\pi)^3\varepsilon_{p_e} }\pder
{\delta a^{(e)}_{\mu}(\vector{p}_e,\vector{x}, t  )}{x_j}
\pder{f^{(\nu)}_0(\vector{q})}{q_j} = 0~, \end{eqnarray}
for neutrinos and
\begin{eqnarray} \label{electron}
&&p_{\mu}\pder{\delta f^{(e)}(\vector{p},\vector{x}, t  )}{x_{\mu}} +
e\delta F_{j\mu}(\vector{x},t)p^{\mu}\pder{f^{(e)}_0(\varepsilon_p)}{p_j}+
eF_{j\mu}^{(0)}p^{\mu}\pder{ \delta f^{(e)}(\vector{p},\vector{x}, t  ) }{p_j}+
\nonumber\\&& + G_Fc_V\sqrt{2}\int
\frac{d^3q}{(2\pi)^3}\frac{(p^{\mu}q_{\mu})}{\varepsilon_q}\pder
{\delta f^{(\nu)}(\vector{q},\vector{x}, t  )}{x_j}
\pder{f^{(e)}_0(\varepsilon_p )}{p_j} +\nonumber\\
&&+ G_Fc_A\sqrt{2} m_e\int
\frac{d^3q}{(2\pi)^3}\frac{(q_{\mu}a^{\mu}_i(\vector{p}))}{\varepsilon_q}
\pder{S^{(0e)}_i(\varepsilon_p )}{p_j}
\pder{\delta f^{(\nu)}(\vector{q},\vector{x}, t)}{x_j} = 0~, \end{eqnarray}
for electrons where $c_V= 2\xi \pm 0.5$, $c_A= \mp 0.5$ are the vector
and axial weak couplings with upper (lower) sign for electron (muon or
tau) neutrinos correspondingly, $\xi=\sin^2\theta_W$ is the
Weinberg parameter; the tensor $F_{jk}^{(0)}=e_{jkl}B_{0l}$ corresponds to
the external magnetic field term.

Here we have introduced
in eq. \pref{neutrino} the 4-vector $a_{\mu}^{(e)}(\vector{p},\vector{x},
t)\equiv a^i_{\mu}(\vector{p})S^i(\vector{p},\vector{x}, t)$ that is the
statistical generalization \cite{Semikoz} of the Pauli-Luba\'nski 4-vector
\cite{BLP}$$a^{\mu}(\vector{p})=(a^{\mu}_i(\vector{p})\zeta_i)=
\left[\frac{\vector{p}\vec{\zeta}}{m_e}; \vec{\zeta} +
\frac{\vector{p}(\vector{p}\vec{\zeta})}{m_e(\varepsilon_p +
m_e)}\right]$$ and has the components \begin{equation} \label{Pauli}
a_{\mu}^{(e)}(\vector{p},\vector{x}, t)=
\left[\frac{\vector{p}\vector{S}^{(e)}(\vector{p},\vector{x}, t)}{m_e};
\vector{S}^{(e)}(\vector{p},\vector{x},t) + \frac{\vector{p}(\vector{p}
\vector{S}^{(e)}(\vector{p},\vector{x}, t) )}{m_e(\varepsilon_e + m_e)}
\right]~. \end{equation}

Note that in
the presence of the external magnetic field $\vector{B}_0$
the neutrino parts of these Vlasov equations differ from the analogous ones
used in \cite{Bingham} due to the inclusion of axial vector
interactions ($\sim c_A$) in the total $\nu$e-scattering amplitude  that
allows us to account for {\it collective interactions of $\nu_{\mu}$
and $\nu_{\tau}$-neutrinos} too for which the vector coupling $c_V$ is
small, $c_V=2\xi - 0.5\approx 0$.

To check general symmetries we temporarily refuse the linearization Eq.
\pref{linear} and write down the RKE for the total
electron spin distribution function $\vector{S}^{(e)}
(\vector{p},\vector{x}, t)$ accounting for the weak interaction of
electrons with neutrinos:

\begin{eqnarray} \label{spin}
&&\frac{\partial S_i^{(e)}(\vector{p},\vector{x}, t) }{\partial t} +
\vector{v} \frac{\partial S_i^{(e)}(\vector{p},\vector{x}, t)
}{\partial \vector{x}} +
eF_{j\mu}(\vector{x},t)\frac{p^{\mu}}{\varepsilon_p}\frac{\partial
S^{(e)}_i(\vector{p},\vector{x}, t )}{\partial p_j}+ \nonumber\\&& +
2\mu_B\left[\frac{E_i(\vector{x},t)(\vector{v}
\vector{S}^{(e)}(\vector{p},\vector{x}, t)) -
v_i(\vector{E}(\vector{x},t)\vector{S}^{(e)}(\vector{p},\vector{x}, t
))}{1 + \gamma} + \frac{[\vector{S}^{(e)}(\vector{p},\vector{x}, t )\times
\vector{B}(\vector{x},t)]}{\gamma} \right] +\nonumber\\&&+
 \frac{2G_F\sqrt{2} c_Am_e}{\varepsilon_p}e_{ikl} \int
\frac{d^3q}{(2\pi)^3}\frac{(q_{\mu}a^{\mu}_k(\vector{p}))}{\varepsilon_q}
f^{(\nu)}(\vector{q},\vector{x}, t)S_l^{(e)}(\vector{p},\vector{x}, t)
+\nonumber\\&&+\frac{G_F\sqrt{2} c_Am_e}{\varepsilon_p}\int
\frac{d^3q}{(2\pi)^3}\frac{(q_{\mu}a^{\mu}_i(\vector{p}))}{\varepsilon_q}
\frac{\partial f^{(e)}(\vector{p},\vector{x}, t )}{\partial p_j}
\frac{\partial f^{(\nu)}(\vector{q},\vector{x}, t)}{\partial x_j}
+\nonumber\\&&+\frac{G_F\sqrt{2} c_V}{\varepsilon_p}\int
\frac{d^3q}{(2\pi)^3}\frac{(p^{\mu}q_{\mu})}{\varepsilon_q}
\frac{\partial S^{(e)}_i(\vector{p},\vector{x}, t )}{\partial p_j}
\frac{\partial f^{(\nu)}(\vector{q},\vector{x}, t)}{\partial x_j}+\nonumber\\&&+
\frac{G_F\sqrt{2}c_V}{\varepsilon_pm_e}[e_{kmn}\int\frac{d^3q}{(2\pi)^3}
\frac{p_mq_k}{\varepsilon_q}\left(v_s\frac{\partial f^{(\nu)}
(\vector{q},\vector{x},t)}{\partial x_s}\right) -\nonumber\\&&-
e_{\mu m\rho \sigma}\int \frac{d^3q}{(2\pi)^3}
\frac{q^{\mu}p^{\rho}a^{\sigma}_n(\vector{p})}{\varepsilon_q}
\frac{\partial f^{(\nu)}(\vector{q}, \vector{x},t)}{\partial x_m}]
e_{ijn}S_j^{(e)}(\vector{p}, \vector{x},t)
= 0~.
\end{eqnarray}
This is the relativistic generalization in SM of the kinetic
equation for spin waves in nonferromagnetic metals \cite{Silin}. But it
should be kept in mind that we consider here the Fermi gas of
free electrons in contrast with the quasi-particle approach for metals and
also neglected  exchange interactions (neglecting exchange Feynman
diagrams for the ee-scattering means here that the long-range forces are
dominant).

The system of RKE's is completed by the Maxwell
equations for the electromagnetic field $F_{\mu \nu}(\vector{x},t)=
\delta F_{\mu \nu}(\vector{x},t) + F_{\mu \nu}^{(0)}$ accounting for the
spin wave contribution, $$- \frac{1}{c}\pder{\vector{B}(\vector{x},t)}{t}=
[\nabla\times \vector{E}(\vector{x},t)]~,$$ and \begin{eqnarray}
\label{Maxwell} &&[\nabla\times [\nabla\times \vector{E}(\vector{x},t)]] +
\frac{1}{c^2}\frac{\partial^2\vector{E}(\vector{x},t)}{\partial t^2} =
-\frac{4\pi}{c^2}\pder{}{t}\Bigl (e\int \frac{d^3p}{(2\pi)^3}\vector{v}
f^{(e)}(\vector{p},\vector{x}, t ) + \nonumber\\&&+\mu_B\left[
\int \frac{d^3p}{(2\pi)^3}\frac{[\nabla \times
\vector{S}^{(e)}(\vector{p},\vector{x}, t)]}{\gamma} - \int
\frac{d^3p}{(2\pi)^3}(\vector{v}\nabla)\frac{[\vector{v}\times
\vector{S}^{(e)}(\vector{p},\vector{x}, t)]}{\gamma +1}\right] \Bigr)~.
\end{eqnarray}
In eqs. \pref{neutrino}--\pref{Maxwell} $e= - \mid e\mid$ is the
electric charge of the electron; $G_F$ is the Fermi constant; $\mu_B=
e\hbar/2m_ec$ is the Bohr magneton; $\vector{v}= \vector{p}/\varepsilon_p$
is the electron velocity;
the latin indices run $i,j=1,2,3$ and the greek ones are $\mu=0,1,2,3$ for
scalars written in the Feynman metrics, $A_{\mu}B^{\mu}= A_0B_0 - A_iB_i$.

In the non-relativistic (NR) limit $\mid\vector{v}\mid\ll 1$ ($\gamma \to
1$) the last term drops out and eq.  \pref{Maxwell} coincides with the
Maxwell equation in a magnetized medium:  \begin{equation}
\label{Maxwell2}
[\nabla \times (\vector{B} - 4\pi \vector{M})] = 4\pi \vector{j} +
\frac{\partial \vector{E}}{\partial t}~, \end{equation}
where $\vector{M}(\vector{x},t)= \mid \mu_B\mid \int
d^3p\vector{S}(\vector{p},\vector{x},t)/(2\pi )^3$ is the magnetization
density of NR plasma.

Multiplying eq. \pref{spin} by the energy $\varepsilon_p$
and the spin distribution function $S_i(\vector{p},\vector{x}, t)$ we
obtain the covariant RKE for the Lorentz-invariant product
$a_{\mu}(\vector{p},\vector{x}, t) a^{\mu}(\vector{p},\vector{x}, t) = -
(\vector{S}^{(e)}(\vector{p},\vector{x}, t) )^2$ where
the 4-vector $a_{\mu}^{(e)}(\vector{p},\vector{x})$ is given by
\pref{Pauli}.

Note that for the uniform electron beam
($S_i^{(e)}(\vector{p},t)= S_i(t)(2\pi )^3n_{e0}\delta^{(3)}(\vector{p} -
\vector{p}_0)$)\\ omitting the neutrino term and integrating the spin RKE
\pref{spin} over $d^3p$  we obtain the one-particle spin evolution
equation $$\frac{d\vector{S}(t)}{dt} = \frac{2\mu_B}{\gamma_0}\left
[\vector{S}\times \vector{B}\right ] + \frac{2\mu_B}{\gamma_0 +
1}\left[\vector{S}\times \Bigl[\vector{E}\times
\vector{v}_0\Bigr]\right]~,$$ that turns out to be exactly the
Bargman-Mishel-Telegdi equation for the electron spin motion with the {\it
normal} magnetic moment $\mu_B$ \cite{BLP}. Here
$\vector{v}_0=\vector{p}/\varepsilon_{p_0}$ is the velocity of the
electron beam, $\gamma_0=\varepsilon_{p_0}/m_e$ is its $\gamma$-factor.

In general, it is possible to
generalize eq.  \pref{spin} adding the electromagnetic scattering of
electrons through {\it anomalous} magnetic moment (Schwinger correction
$\mu' = (\alpha/2\pi)\mu_B$) that could lead to additional terms similar
to electromagnetic terms in the neutrino spin RKE \cite{Semikoz2}.

Below we consider some important particular cases of the master equations
checking their consistency with the known results
\cite{Bingham,Bento,Brizard}.  \vskip 0.3cm \subsection{Lepton current
conservation $\partial j^{(a)}_{\mu}(\vector{x},t)/\partial
x_{\mu}=0$} \vskip 0.3cm

Let us rewrite eq. (\ref{neutrino}) as the
classical equation \begin{equation}\label{neutrino3}\frac{\partial
\delta f^{(\nu)}(\vector{q}, \vector{x},t)}{\partial t} +
\dot{\vector{x}}\frac{\partial  \delta f^{(\nu)}(\vector{q},\vector{x},t)}
{\partial \vector{x}} + \dot{\vector{q}}
\frac{\partial f^{(\nu)}_0(\vector{q})}
{\partial \vector{q}} =0~,\end{equation}
where $\dot{\vector{x}}=\vector{n}= \vector{q}/q$ is the velocity of
massless neutrino, the derivative $\dot{\vector{q}}$ is given
by \begin{eqnarray} \label{qforce} \dot{\vector{q}}= +
&&\nabla\left[G_F\sqrt{2}c_V\delta n^{(e)}(\vector{x},t)
+ \frac{G_F}{\sqrt{2}}c_A\delta A_0(\vector{x},t)\right] - \nonumber\\&&-
G_F\left[
\sqrt{2}c_V\nabla [\vector{n}\delta \vector{j}^{(e)}(\vector{x},t)] +
\frac{c_A}{\sqrt{2}}\nabla[\vector{n}
\delta \vector{A}^{(e)}(\vector{x},t)] \right]~.  \end{eqnarray} Here  $
\delta j_{\mu}^{(e)}(\vector{x},t)= (\delta n^{(e)}(\vector{x},t);
\delta \vector{j}^{(e)}(\vector{x},t) )=\int
(d^3p/(2\pi)^3)(p_{\mu}/\varepsilon_p) \delta f^{(e)}(\vector{p},
\vector{x},t)$ is the four-vector of the electron current density
perturbation ;\\ $\delta A_{\mu}^{(e)}(\vector{x},t)= m_e\int
(d^3p/(2\pi)^3)\delta a_{\mu}^{(e)} (\vector{p},\vector{x}, t
)/\varepsilon_p$ is the axial four-vector of the spin density
perturbation\footnote{In NR plasma the spin density is given by the
3-vector component only, $\delta A_{\mu}(\vector{x},t) = (0,
\delta \vector{S}(\vector{x},t))$~.}; the four-vector spin distribution
$\delta a_{\mu}^{(e)} (\vector{p},\vector{x}, t )$ is given by
(\ref{Pauli});$\nabla\equiv \partial_i= \partial /\partial x^i= -
\partial^i $.

Analogously the electron
RKE (\ref{electron}) can be rewritten as the quasi-classical one,
\begin{eqnarray}
\label{pRKE}
\frac{\partial \delta f^{(e)}(\vector{p},\vector{x},t)}{\partial t} + \dot{\vector{x}}\frac{\partial \delta f^{(e)}(\vector{p},\vector{x},t)}{\partial \vector{x}} + Tr\left(\dot{\hat{\vector{p}}}\frac{\partial \hat{f}^{(0e)}(\varepsilon_p)}{\partial \vector{p}}\right) = 0~,
\end{eqnarray}
where the particle number distribution function is obtained via the summing over spin quantum numbers,
$\delta f^{(e)}(\vector{p},\vector{x},t) = Tr(\delta \hat{f}^{(e)}(\vector{p},\vector{x},t)$; $\dot{\vector{x}}= \vector{v}= \vector{p}/\varepsilon_p$ is the electron velocity;
\begin{eqnarray}\label{pdot}\dot{\hat{\vector{p}}}=
&&\left(e\Bigl(\delta \vector{E}(\vector{x},t) + [\vector{v}\times
\vector{B}(\vector{x},t)]\Bigr) +
G_F\sqrt{2}c_V\Bigl[\frac{p^{\mu}}{\varepsilon_p}\nabla \delta
j^{(\nu)}_{\mu}(\vector{x},t)\Bigr]\right)\delta_{\lambda \lambda^{'}}
+\nonumber\\&&+ G_F\sqrt{2}c_A\frac{m_e}{\varepsilon_p}\nabla \delta
j_{\mu}^{(\nu)}(\vector{x},t)a^{\mu}_k(\vector{p})(\vector{\sigma}_k)_{\lambda
\lambda^{'}}~ \end{eqnarray} is the force matrix which accounts
for the Lorentz force with electromagnetic fields $\delta \vector{E}$,
$\vector{B}= \vector{B}_0 + \delta \vector{B}$ as well as the weak
interaction terms $\sim G_F$ \footnote{ Such tems can be also obtained
from the weak interaction Hamiltonian for a probe electron moving in a
neutrino medium $$\hat{H}_{weak} = - \hat{L}_{weak}=
G_F\sqrt{2}\frac{\bar{u}_{\lambda}(p)
\gamma^{\mu}u_{\lambda^{'}}(p)}{2\varepsilon_p}
\delta j^{(\nu)}_{\mu}(\vector{x},t) +
G_F\sqrt{2}c_A\frac{\bar{u}_{\lambda}(p)\gamma^{\mu}
\gamma_5u_{\lambda^{'}}(p)}{2\varepsilon_p}
\delta j^{(\nu)}_{\mu}(\vector{x},t)~.$$};
\begin{equation}
\label{Fermi1}
\hat{f}^{(0e)}(\varepsilon_p) =
\frac{\delta_{\lambda^{'}\lambda}}{2}f^{0e}(\varepsilon_p) +
\frac{(\vec{\sigma}\hat{\vec{b}}^{(0)})_{\lambda^{'}\lambda}}{2}S^{(0e)}(\varepsilon_p)
\end{equation}
is the equilibrium density matrix as given in \pref{background} with the
ort $\hat{\vector{b}}^{(0)}= \vector{B}_0/B_0$ separated from the isotropic
distribution function $S^{(0e)}(\varepsilon_p)= - (\mid \mu_B\mid
B_0/\gamma)\mbox{\rm d}f^{(0e)}/\mbox{\rm d}\varepsilon_p$ given by
\pref{polar}.

Substituting \pref{pdot}, \pref{Fermi1} into \pref{pRKE}
one can easily check that the electron RKE \pref{electron} takes the form
\begin{eqnarray}\label{electron1}&&\frac{\partial \delta
f^{(e)}(\vector{p}, \vector{x},t)}{\partial t} + \vector{v}\frac{\partial
\delta f^{(e)}(\vector{p}, \vector{x},t)}{\partial \vector{x}}+ e\delta
F_{j\mu}(\vector{x},t)\frac{p^{\mu}}{\varepsilon_p}
\frac{\partial f^{(e)}_0(\varepsilon_p)}{\partial p_j}+
eF_{j\mu}^{(0)}\frac{p^{\mu}}{\varepsilon_p}
\pder{ \delta f^{(e)}(\vector{p},\vector{x}, t  ) }{p_j} + \nonumber\\&&
+ G_F\sqrt{2}c_V\left(\frac{\partial
\delta n^{(\nu)}(\vector{x},t)}{\partial x_j} - v_k\frac{\partial
\delta j^{(\nu)}_k(\vector{x},t)}{\partial x_j} \right)
\frac{\partial f^{(e)}_0(\varepsilon_p)}{\partial p_j} + \nonumber\\&&
+G_F\sqrt{2}c_Am_e\left(\frac{\partial
\delta n^{(\nu)}(\vector{x},t)}{\partial
x_j}\frac{a_0(\vector{p})}{\varepsilon_p} -
\frac{a_k(\vector{p})}{\varepsilon_p}\frac{\partial \delta
j^{(\nu)}_k(\vector{x},t)}{\partial x_j} \right) \frac{\partial
S^{(0e)}(\varepsilon_p)}{\partial p_j}= 0~,
\end{eqnarray}
where we introduced the unit polarization
four-vector $a^{\mu}(\vector{p})= a^{\mu}_i(\vector{p}) \hat{b}^{(0)}_i $,
$a_{\mu}a^{\mu} = -1$.

On first glance both the neutrino (\ref{neutrino3}) and electron
(\ref{pRKE}) RKE's {\it do not obey the lepton number conservation law},
$\partial
j^{(a)}_{\mu}(\vector{x},t)/\partial x_{\mu}\neq 0$, (due to the last
terms in (\ref{qforce}) and the last three terms in (\ref{electron1})
correspondingly).

This non-conservation is true for the gauge non-invariant Wigner
distributions $f^{(a)}(\vector{p},\vector{x},t)\equiv\tilde{f}^{(a)}(\vector{p},\vector{x},t)=\int
e^{i\vector{p}\vector{y}}d^3y\tilde{f}^{(a)}(\vector{x}-\vector{y}/2,
\vector{x}+ \vector{y}/2,t)$ we used above, for which the distribution
function in the coordinate representation (within integrand), $$
\tilde{f}^{(a)}(\vector{x}_1,\vector{x}_2,t)=
Tr\left(\hat{\rho}(t)\hat{\Psi}^{(a)+}(\vector{x}_2)
\hat{\Psi}^{(a)}(\vector{x}_1)\right)~,
$$
is not invariant with respect to gauge transformation of the
corresponding interaction fields.

For example, in QED plasma the kinetic
equation for the gauge non-invariant distribution of charged particles
$\tilde{f}^{(e)}(\vector{x}_1,\vector{x}_2,t)$ derived from the quantum
Liouville equation by the same Bogolyubov method {\it does not obey the
electric current conservation law} since the force term depends on the
electromagnetic potentials $A_{\mu}(\vector{x},t)$ which do not enter as
combinations expressed via field strengths, $\vector{E}$, $\vector{B}$
\cite{Peletminsky}.

The recasting of such RKE for the gauge-invariant Wigner distribution
\begin{equation}\label{chargeWigner}
f^{(e)}(\vector{p},\vector{x},t)=\tilde{f}^{(e)}(\vector{p}+
e\vector{A}(\vector{x},t) , \vector{x},t)=\int
d^3ye^{i\vector{p}\vector{y}}f^{(e)}(\vector{x}-
\vector{y}/2,\vector{x} + \vector{y}/2,t)~,\end{equation}
where the gauge invariant
distribution in the coordinate representation
$f^{(e)}(\vector{x}_1,\vector{x}_2,t)$ is connected with the gauge
non-invariant one,\\ $\tilde{f}^{(e)}(\vector{x}_1,\vector{x}_2,t)=
Tr\left(\hat{\rho}(t)\hat{\Psi}^{(e)+}(\vector{x}_2)\hat{\Psi}^{(e)}(
\vector{x}_1)\right)$,
by the important phase factor \cite{Fujita}:
\begin{equation}
\label{phase}
f^{(e)}(\vector{x}_1,\vector{x}_2,t)=\exp
\left[ie(\vector{x}_2-\vector{x}_1)
\int_0^1d\xi\vector{A}\Bigl(\vector{x}_2
+ \xi(\vector{x}_1-
\vector{x}_2),t\Bigr)\right]\tilde{f}^{(e)}(\vector{x}_1,\vector{x}_2,t)~,
\end{equation}
allows to obtain the usual Lorentz force
term in the standard Boltzman equation for charged particles
\cite{Peletminsky}:
$$\frac{\partial
f^{(e)}(\vector{p},\vector{x},t)}{\partial t}+
\vector{v}\frac{\partial f^{(e)}(\vector{p},\vector{x},t)}{\partial
\vector{x}} + e\left(\vector{E} + [\vector{v}\times
\vector{B}]\right)\frac{\partial
f^{(e)}(\vector{p},\vector{x},t)}{\partial \vector{p}}=0~,$$
for which, of course, the electric current $j_{\mu}^{(e)}(\vector{x},t)=
\int d^3p(p_{\mu}/\varepsilon_p)f^{(e)}(\vector{p},\vector{x},t)/(2\pi)^3$
is conserved, $\partial j_{\mu}^{(e)}/\partial x_{\mu}=0$.

Such conservation is stipulated by the presence of the phase factor in the
distribution (\ref{phase}) which in turn is invariant under the standard
gauge transformation (with an arbitrary gauge function $\chi
(\vector{x},t)$ obeying the d'Alambert equation), \begin{eqnarray}
\label{standard} &&\hat{\Psi}^{(e)}(\vector{x}_1)\to
e^{-ie\chi(\vector{x}_1,t)} \hat{\Psi}^{(e)}(\vector{x}_1)~,\nonumber\\&&
\hat{\Psi}^{(e)+}(\vector{x}_2)\to e^{+ie\chi(\vector{x}_2,t)}
\hat{\Psi}^{(e)}(\vector{x}_2)~,\nonumber\\&&
\vector{A}\left(\vector{x}_2 + \xi(\vector{x}_1- \vector{x}_2),t\right)\to
\vector{A}\left(\vector{x}_2 + \xi(\vector{x}_1- \vector{x}_2),t\right) -\nonumber\\&&-
\frac{\partial \chi \left(\vector{x}_2 + \xi(\vector{x}_1-
\vector{x}_2),t\right)}{\partial \vector{x}_2}~,  \end{eqnarray}
or, equivalently, this arbitrary phase $\chi (\vector{x},t)$ cancels in
(\ref{phase}).  Such invariance is crucial for macroscopic physics since
it provides the physical sense of the Wigner function (\ref{chargeWigner})
and the conservation of the macroscopic electric current.

Deriving electron RKE (\ref{pRKE}) we used such gauge transformation while
for the weak interaction terms we did not.

Thus, a recipe of the gauge invariance restoration in RKE's
(\ref{neutrino3}), (\ref{pRKE}) should be similar to the change of
arguments in (\ref{chargeWigner}): we should substitute the generalized
(congugate) neutrino momentum, \begin{equation} \label{canonical}
\vector{Q}= \vector{q} + G_F\sqrt{2}c_V \vector{j}^{(e)}(\vector{x},t) +
\frac{G_F}{\sqrt{2}}c_A \vector{A}^{(e)}(\vector{x},t)~, \end{equation}
for the neutrino gauge invariant Wigner distribution
\begin{equation}\label{neutrinoWigner}
f^{(\nu)}(\vector{q},\vector{x},t)=
\tilde{f}^{(\nu)}(\vector{Q},\vector{x},t)~,\end{equation} and
the matrix of the generalized momentum
\begin{equation}\label{Pmomentum} \vector{\hat{P}}= [\vector{p} +
G_F\sqrt{2}c_V \delta \vector{j}^{(\nu)}(\vector{x},t)]\delta_{\lambda
\lambda^{'}} - G_F\sqrt{2}c_A\delta n^{(\nu)}(\vector{x},t)
(\vec{\sigma})_{\lambda \lambda^{'}}~,
\end{equation}
for the electron gauge
invariant Wigner distribution
\begin{equation}\label{electronWigner}
f^{(e)}(\vector{p},\vector{x},t)=
       \tilde{f}^{(e)}(\vector{P},\vector{x},t)~,\end{equation}
where $\vector{P}= 2^{-1}Tr~\hat{\vector{P}}$.

The  RKE's for gauge invariant
(\ref{neutrinoWigner}),(\ref{electronWigner}) are obtained from
(\ref{neutrino3}), (\ref{pRKE}) as $$ \frac{\partial
\delta \tilde{f}^{(\nu)}(\vector{Q}, \vector{x},t)}{\partial t} +
\dot{\vector{x}}\frac{\delta \partial
\tilde{f}^{(\nu)}(\vector{Q},\vector{x},t)} {\partial \vector{x}} +
\dot{\vector{Q}} \frac{\partial
\tilde{f}^{(\nu)}_0(\vector{Q})} {\partial \vector{Q}} =0~, $$ and $$
\frac{\partial \delta \tilde{f}^{(e)}(\vector{P},\vector{x},t)}{\partial
t} + \dot{\vector{x}} \frac{\partial \delta
\tilde{f}^{(e)}(\vector{P},\vector{x},t)}{\partial \vector{x}} +
Tr~ \left[\dot{\hat{\vector{P}}}\frac{\partial
\hat{f}^{(e)}_0(\varepsilon_P)}{\partial \vector{P}}\right] = 0~, $$
where generalized momenta $\vector{Q}$, $\hat{\vector{P}}$ are given by
Eqs.  (\ref{canonical}) and (\ref{Pmomentum}) correspondingly; the
background distributions $\tilde{f}^{(a)}_0$ can be changed to
$f^{(\nu)}_0(\vector{q})$ for neutrinos and $\hat{f}^{(e)}_0(\varepsilon_p)$
(\ref{Fermi1}) for electrons since we retain the lowest order $\sim
G_F$ in the corresponding weak interaction terms.

Substituting in last RKE's the momenta (\ref{canonical}) and
(\ref{Pmomentum}), for which we take into account both the derivatives
(\ref{qforce}), (\ref{pdot}) and the total time derivatives
$\rm{d}\delta \vector{j}^{(e)}(\vector{x},t)/\rm{dt}=\partial
\delta \vector{j}^{(e)}(\vector{x},t)/\partial t +
(\vector{n}\nabla)\delta \vector{j}^{(e)}(\vector{x},t)$,
$\rm{d}\delta \vector{A}^{(e)}(\vector{x},t)/\rm{dt}=\partial \delta
\vector{A}^{(e)}(\vector{x},t)/\partial t +
(\vector{n}\nabla)\delta \vector{A}^{(e)}(\vector{x},t)$ in the neutrino
RKE, $\rm{d}\delta j^{(\nu)}_{\mu}(\vector{x},t)/\rm{dt}=\partial \delta
j^{(\nu)}_{\mu}(\vector{x},t)/\partial t + (\vector{v}\nabla)\delta
j^{(\nu)}_{\mu}(\vector{x},t)$ in
the electron one, and then using the identities $$
(\vector{n}\nabla)\delta \vector{j}^{(e)}(\vector{x},t) -
\nabla(\vector{n}\delta \vector{j}^{(e)}(\vector{x},t))\equiv -
[\vector{n}\times \nabla\times \delta \vector{j}^{(e)}(\vector{x},t)]~,$$
$$ (\vector{n}\nabla)\delta \vector{A}^{(e)}(\vector{x},t) -
\nabla(\vector{n}\delta \vector{A}^{(e)}(\vector{x},t))\equiv -
[\vector{n}\times \nabla\times \delta \vector{A}^{(e)}(\vector{x},t)]~,
$$ $$ (\vector{v}\nabla)\delta \vector{j}^{(\nu)}(\vector{x},t) -
\nabla(\vector{v}\delta \vector{j}^{(\nu)}(\vector{x},t))\equiv -
[\vector{v}\times \nabla\times \delta
\vector{j}^{(\nu)}(\vector{x},t)]~,$$ $$ (\vector{v}\nabla) \delta
n^{(\nu)}(\vector{x},t)\hat{\vector{b}}^{(0)}- \nabla
(\vector{v}\hat{\vector{b}}^{(0)}\delta n^{(\nu)}(\vector{x},t))\equiv -
(\vector{v}\times \nabla\times \hat{\vector{b}}^{(0)}\delta
n^{(\nu)}(\vector{x},t))~, $$
we arrive to the final forms of the
lepton RKE's.

Namely, accounting for the definition of the neutrino gauge
invariant distribution (\ref{neutrinoWigner}) the neutrino RKE
(\ref{neutrino3}) which is equivalent to the master (\ref{neutrino}) takes
finally in the Vlasov approximation the form \begin{eqnarray}
\label{neutrino1}\frac{\partial \delta f^{(\nu)}(\vector{q},\vector{x}, t
)}{\partial t} + &&\vector{n}\frac{\partial
\delta f^{(\nu)}(\vector{q},\vector{x}, t )}{\partial \vector{x}} + \delta
F_{j\mu}^{(V)}(\vector{x},t)\frac{q^{\mu}}{\varepsilon_q}\frac{\partial
f^{(\nu)}_0(\vector{q})}{\partial q_j} + \nonumber\\&&+
\delta F_{j\mu}^{(A)}(\vector{x},t)\frac{q^{\mu}}{\varepsilon_q}
\frac{\partial f^{(\nu)}_0(\vector{q})}{\partial q_j}=0~,
\end{eqnarray}
where the
antisymmetric tensors $\delta F_{jk}^{(V,A)}(\vector{x},t)$ entering
effective Lorentz force terms are given by the vector and axial vector
currents correspondigly, \begin{eqnarray} \label{tensors}
&&\delta F_{j0}^{(V)}(\vector{x},t)/G_F\sqrt{2}c_V= -\nabla_j
\delta n^{(e)}(\vector{x},t) - \frac{\partial
\delta j^{(e)}_j(\vector{x},t)}{\partial
t}~,\nonumber\\&&\delta F_{jk}^{(V)}(\vector{x},t)/G_F\sqrt{2}c_V=
e_{jkl}(\nabla\times
\delta \vector{j}^{(e)}(\vector{x},t))_l~,\nonumber\\&&
\sqrt{2}\delta F_{j0}^{(A)}(\vector{x},t)/G_Fc_A= -\nabla_j
\delta A^{(e)}_0(\vector{x},t) - \frac{\partial
\delta A^{(e)}_j(\vector{x},t)}{\partial
t}~,\nonumber\\&&\sqrt{2}\delta F_{jk}^{(A)}(\vector{x},t)/G_Fc_A=
e_{jkl}(\nabla\times
\delta \vector{A}^{(e)}(\vector{x},t))_l~;
\end{eqnarray}
and $\partial /\partial \vector{Q}=\partial/\partial \vector{q}$ from the
relation (\ref{canonical}).

First three terms
in RKE (\ref{neutrino1}) were derived in \cite{Bingham,Bento}.
In the paper \cite{Brizard} analogous vector coupling terms
($\sim c_V$ for electron neutrinos) were obtained in cold hydrodynamics.

Analogously for the electron gauge invariant distribution
(\ref{electronWigner}) we obtain finally from \pref{electron1} (equivalent
to the master equation (\ref{electron})) the electron RKE
\begin{eqnarray}\label{electron2}&&\frac{\partial \delta
f^{(e)}(\vector{p}, \vector{x},t)}{\partial t} + \vector{v}\frac{\partial
\delta f^{(e)}(\vector{p}, \vector{x},t)}{\partial \vector{x}}+ e\delta
F_{j\mu}(\vector{x},t)\frac{p^{\mu}}{\varepsilon_p}
\frac{\partial f^{(e)}_0(\varepsilon_p)}{\partial p_j}+
eF_{j\mu}^{(0)}\frac{p^{\mu}}{\varepsilon_p}
\pder{ \delta f^{(e)}(\vector{p},\vector{x}, t  ) }{p_j} + \nonumber\\&&
 + F_{j\mu}^{(weak)}(\vector{x},t)\frac{
p^{\mu}}{\varepsilon_p}\frac{\partial f^{(e)}_0(\varepsilon_p)}{\partial
p_j}
- G_F\sqrt{2}c_A\Bigl[\frac{\partial
\delta n^{(\nu)}(\vector{x},t)\hat{\vector{b}}^{(0)}}{\partial t}
- (\vector{v}\times \nabla\times \delta
n^{(\nu)}(\vector{x},t)\hat{\vector{b}}^{(0)}) + \nonumber\\&& +
\frac{m_e}{\varepsilon_p}\nabla (\vector{a}(\vector{p}) \delta
\vector{j}^{(e)}(\vector{x},t)) \Bigr] \frac{\partial
S^{(0e)}(\varepsilon_p)}{\partial p_j}= 0~,
\end{eqnarray}
where the weak vector term ($\sim c_V$) has the Lorentz structure
 with the tensor components \cite{Bingham,Bento}\begin{eqnarray}
\label{electrontensor} &&F_{j0}^{(weak)}(\vector{x},t)/G_F\sqrt{2}c_V=
-\nabla_j\delta n^{(\nu)}(\vector{x},t) - \frac{\partial \delta
j^{(\nu)}_j(\vector{x},t)}{\partial
t}~,\nonumber\\&&F_{jk}^{(weak)}(\vector{x},t)/G_F\sqrt{2}c_V=
e_{jkl}(\nabla\times \delta
\vector{j}^{(\nu)}(\vector{x},t))_l~;
\end{eqnarray}
$\partial/\partial \vector{P}=\partial/\partial \vector{p}$ from the
relation (\ref{Pmomentum});
and the axial vector term does not contribute to the continuity
equation since the last term with the 3-vector component of the
four-vector $a_{\mu}(\vector{p})$ introduced after \pref{electron1} ,
$$\vector{a}(\vector{p}) =\hat{\vector{b}}^{(0)} +
\frac{\vector{p}(\vector{p}\cdot \hat{\vector{b}}^{(0)}
)}{m_e(\varepsilon_p + m_e)}~,$$ vanishes under integration $\int
d^3p(a_k(\vector{p})/\varepsilon_p)\partial S^{(0e)}(\varepsilon_p)/
\partial
p_j$ due to the isotropy of the background and the {\it
odd} power of momenta $\vector{p}$ under the integral.

Obviously, RKE's (\ref{neutrino1}), (\ref{electron2}) obey the lepton
current conservation law
\begin{equation}
\label{law}
\frac{\partial
j^{(a)}_{\mu}(\vector{x},t)}{\partial x_{\mu}}=0,
\end{equation}
or the lepton currents are conserved in a magnetized plasma.
\vskip 0.3cm \subsection{Dispersion equation for
lepton density perturbations in isotropic plasma} \vskip 0.3cm
Here checking master equations (\ref{neutrino}), (\ref{electron})
against known results for an isotropic plasma in
the absence of an external magnetic field, $B_0=0$, hence neglecting spin
waves, we show that accounting for
the lepton current conservation (\ref{law}) such observables like the
spectra of lepton density perturbations do not depend whether we apply
initial RKE's for gauge non-invariant distribution functions
(\ref{neutrino}), (\ref{electron}), or same equations written in the
completed forms, (\ref{neutrino1}),(\ref{electron2}).

The situation is similar to the case
of standard plasma where the initial RKE for the gauge
non-invariant Wigner distribution
$\tilde{f}^{(e)}(\vector{p},\vector{x},t)$ is often more suitable for the
obtaining of concrete results than Boltzman equation with the Lorentz
force \cite{Peletminsky}. Nevertheless, the electric current should be
written in the gauge invariant form obeying (\ref{law}).

As well as authors \cite{Bingham,Bento} we
consider the electron plasma waves (EPW) driven by the intense neutrino
flux, neglecting also the transverse wave contribution, i.e. retaining
only the electric field in the Lorentz force for electrons.

Using the Fouirer representation in the RKE's \pref{neutrino},
\pref{electron} one can easily obtain the algebraic system
 for the current perturbations\\
$\delta j^{(\nu)}_{\mu}(\omega, \vector{k})=\int d^3q(q_{\mu}/q)\delta
f^{(\nu)}(\vector{q}, \vector{k},\omega)/(2\pi)^3 $, \\
$\delta j^{(e)}_{\mu}(\omega, \vector{k})=\int
d^3p(p_{\mu}/\varepsilon_p)\delta f^{(e)}(\vector{p},
\vector{k},\omega)/(2\pi)^3 $, \begin{eqnarray} \label{system} &&\delta
n^{(\nu)}(\omega, \vector{k}) + G_F\sqrt{2}c_V\Bigl [\delta n^{(e)}(\omega,
\vector{k})\int\frac{d^3q}{(2\pi)^3}\frac{k_j\partial
f^{(\nu)}_0(q)/\partial q_j}{\omega - \vector{k}\vector{n}} -\nonumber\\&&
-\delta j^{(e)}_k (\omega,
\vector{k})k_j \int\frac{d^3q}{(2\pi)^3}\frac{n_k \partial
f^{(\nu)}_0(q)/\partial q_j}{\omega - \vector{k}\vector{n}}\Bigr ]
=0~,\nonumber\\&&
\delta n^{(e)}(\omega, \vector{k})\Bigl [1 + \chi_e(\omega,
\vector{k})\Bigr ] + G_F\sqrt{2}c_V\Bigl
[\delta n^{(\nu)}(\omega,
\vector{k})\int\frac{d^3p}{(2\pi)^3}\frac{k_j\partial
f^{(e)}_0(\varepsilon_p)/\partial p_j}{\omega - \vector{k}\vector{v}}
-\nonumber\\&& -\delta j^{(\nu)}_k (\omega,
\vector{k})k_j \int\frac{d^3p}{(2\pi)^3}\frac{v_k \partial
f^{(e)}_0(\varepsilon_p )/\partial p_j}{\omega -
\vector{k}\vector{v}}\Bigr ] =0 ~,
\end{eqnarray}
where $\vector{n}= \vector{q}/q$ is the velocity of the massless neutrino
and the term $\chi_e(\omega, \vector{k})$ is connected with the
longitudinal permittivity, \begin{equation}\label{longitudinal}
\chi_e(\omega, \vector{k})=\varepsilon_l(\omega, \vector{k}) - 1 =
\frac{4\pi e^2}{k^2}\int \frac{d^3p}{(2\pi)^3}\frac{k_j \partial
f^{(e)}_0(\varepsilon_p )/\partial p_j}{\omega - \vector{k}\vector{v}}~.
\end{equation}
Note that starting instead of \pref{neutrino}, \pref{electron} from the
corresponding RKE's \pref{neutrino1}, \pref{electron2} derived in the
previous subsection for the gauge invariant Wigner distribution functions
$f^{(\nu)}(\vector{Q},\vector{x},t)$, $f^{(e)}(\vector{P}, \vector{x},t)$
we obtain the same algebraic system \pref{system} omitting
axial terms ($\sim c_A$) for $\vector{B}_0=0$.

Really additional terms coming from the total time derivatives,
$$\frac{\partial \delta \vector{j}^{(e)}(\vector{x},t)}{\partial t} +
(\vector{n}\nabla) \delta \vector{j}^{(e)}(\vector{x},t)$$ for
$\dot{\vector{Q}}$ via the second term in (\ref{canonical}), and
$$\frac{\partial \delta \vector{j}^{(\nu)}(\vector{x},t)}{\partial t} +
(\vector{v}\nabla) \delta\vector{j}^{(\nu)}(\vector{x},t)$$ for
$\dot{\vector{P}}$ via the second term in \pref{Pmomentum}, are
proportional in the Fourier representation to the $\check{C}$erenkov
factors $\omega - \vector{k}\vector{n}$, $\omega - \vector{k}\vector{v}$
correspondingly that exactly cancels with the poles within integrands in
\pref{system} leading to zero contributions in brackets for both equations
\pref{system}: $\sim \delta j^{(e)}(\omega, \vector{k})\int
(d^3q/(2\pi)^3)\partial f^{(\nu)}_0(q)/\partial q_j = 0$, and\\ $\sim
\delta j^{(\nu)}(\omega, \vector{k})\int (d^3p/(2\pi)^3)\partial
f^{(e)}_0(\varepsilon_p)/\partial p_j = 0$ .

Now using the symmetrical tensor entering the second line in \pref{system}
(and the analogous one in the fourth line)
$$
\int\frac{d^3q}{(2\pi)^3}\frac{n_k \partial
f^{(\nu)}_0(q)/\partial q_j}{\omega - \vector{k}\vector{n}}= A\delta_{kj}
+ B\frac{k_kk_j}{k^2}~, $$
where the scalars $A$, $B$ are given by
$$
A = \frac{1}{2}\left[ \int\frac{d^3q}{(2\pi)^3(\omega - \vector{k}\vector{n})}
\left (n_i\frac{\partial f^{(\nu)}_0(q)}{\partial q_i} - \frac{\omega}{k^2}
k_i\frac{\partial f^{(\nu)}_0(q)}{\partial q_i}\right) \right]
$$
$$
B = \frac{1}{2}\left[ \int\frac{d^3q}{(2\pi)^3(\omega -
\vector{k}\vector{n})} \left (\frac{3\omega}{k^2} k_i\frac{\partial
f^{(\nu)}_0(q)}{\partial q_i}  - n_i\frac{\partial
f^{(\nu)}_0(q)}{\partial q_i} \right) \right]~, $$
we obtain the factor multiplying the electron current $\delta j^{(e)}_k$
in the second line (and the analogous one in the fourth line in
\pref{system}) \begin{equation} \label{factor}
k_j\int\frac{d^3q}{(2\pi)^3}\frac{n_k \partial f^{(\nu)}_0(q)/\partial
q_j}{\omega - \vector{k}\vector{n}}=k_k(A + B)=\frac{k_k\omega}{k^2}
\int\frac{d^3q} {(2\pi)^3(\omega - \vector{k}\vector{n})}
k_i\frac{\partial f^{(\nu)}_0(q)}{\partial q_i}~.  \end{equation}
Substituting eqs. \pref{longitudinal}, \pref{factor}
we rewrite the system \pref{system} as
\begin{eqnarray}
\label{system1}
&&\delta n^{(\nu)}(\omega, \vector{k}) + G_F\sqrt{2}c_V\Bigl [\delta
n^{(e)}(\omega, \vector{k}) - \frac{\omega}{k^2}k_k\delta j^{(e)}_k (\omega,
\vector{k})\Bigr ]\times \nonumber\\&&\times \int\frac{d^3q}
{(2\pi)^3(\omega - \vector{k}\vector{n})} k_i\frac{\partial
f^{(\nu)}_0(q)}{\partial q_i} = 0~,\nonumber\\ &&G_F\sqrt{2}c_V\Bigl
[\delta n^{(\nu)}(\omega, \vector{k}) - \frac{\omega}{k^2}k_k\delta
j^{(\nu)}_k (\omega, \vector{k})\Bigr ] \frac{k^2\chi_e (\omega,
\vector{k})}{4\pi e^2} + \nonumber\\&& + \delta n^{(e)}(\omega,
\vector{k})\Bigl (1 + \chi_e (\omega, \vector{k})\Bigr)= 0~.\end{eqnarray}
Using the lepton current conservation (\ref{law}), $k_k\delta
j^{(a)}_k(\omega, \vector{k})=\omega \delta n^{(a)} (\omega, \vector{k})$,
we obtain from \pref{system1} the dispersion equation in the lowest order
over the Fermi constant ($\sim G_F^2$)\cite{Bingham,Bento} :
\begin{equation} \label{dispersion1} 1 + \chi_e (\omega, \vector{k}) -
\frac{\Delta_{\nu}k^2c_V^2}{\omega_{pe}^2}\left(1 -
\frac{\omega^2}{k^2}\right)^2\chi_e(\omega, \vector{k})\int\frac{d^3q}
{(2\pi)^3(\omega - \vector{k}\vector{n})} k_i\frac{\partial
\hat{f}^{(\nu)}_0(q)}{\partial q_i} = 0~,  \end{equation}
where $\Delta_{\nu}= 2G_F^2n_{e0}n_{\nu 0}/m_e$ is the weak
parameter introduced in \cite{Bingham}; $\hat{f}^{(\nu)}_0 =
f^{(\nu)}_0(q)/n_{\nu 0}$ is the normalized neutrino background
distribution function; $\omega_{pe}=\sqrt{4\pi \alpha
n_{e0}/m_e}$ is the non-relativistic expression for the plasma frequency.

Let us stress the importance of additional terms in
RKE's \pref{neutrino1}, \pref{electron2} discussed after eq.
\pref{longitudinal} which did not contribute to \pref{system1} while they
become important when we claim the lepton current conservation deriving
dispersion equation \pref{dispersion1}.  We can conclude that the master
RKE's \pref{neutrino}, \pref{electron} plus the lepton current
conservation laws provided by corresponding eqs. \pref{neutrino1},
\pref{electron2} lead to the correct issues known in literature, in
particular, to the dispersion equation \pref{dispersion1}.

The prediction of the
streaming instability driven by the neutrino beam in a dense plasma (i.e.,
outside the neutrinosphere of a supernova) \cite{Bingham} was criticized
in \cite{Bento,Melrose,Laming}.  In this section, however, we do not touch
these issues following from eq.  \pref{dispersion1}.

Let us turn to the case of magnetized plasma.
\section{Bloch equation in the presence of neutrino beam}

Substituting the linear decomposition
\pref{linear} into eq.  \pref{spin}, then integrating the latter over
$d^3p$ and multiplying by $\mu_B$ we have generalized Bloch equation for
the magnetization density perturbations $\vector{m}(\vector{x}, t)= \mid
\mu_B\mid\int d^3p \delta \vector{S}(\vector{p},\vector{x}, t)  /(2\pi)^3$
 in a polarized NR electron gas
that is the base of the theory of the
electron paramagnetic resonance in the absence of
weak interactions \cite{Bloch} while in SM with neutrinos such
equation takes the form
\begin{eqnarray} \label{Bloch} \frac{\partial
m_j(\vector{x},t)}{\partial t} + && 2\mu_B \left [ \Bigl
(\vector{m}(\vector{x},t) - \chi_0\vector{b}(\vector{x},t) \Bigr )\times
\vector{B}_0 \right ]_j + \frac{\partial
\Sigma_{kj}(\vector{x},t)}{\partial x_k} +\nonumber\\&& +
\frac{G_F\sqrt{2} c_A\mu_Bn_{0e}}{m_e}\int
\frac{d^3q}{(2\pi)^3}\frac{\partial \delta
f^{(\nu)}(\vector{q},\vector{x},t)}{\partial x_j} = 0~.
\end{eqnarray}
Here $\chi_0= - 2\mu_B^2\int d^3p(df^{(0)}/d\varepsilon_p)/(2\pi)^3$ is
the static susceptibility of the polarized electron gas \footnote{
The static susceptibility is small in
a degenerate electron gas , $\chi_0= \alpha v_{F_e}/4\pi^2\ll 1$ ,
in contrast with the varying one $\chi (t)$
(see below eq \pref{resonance}). This is the reason why the static
magnetic induction $B_0= (1 + 4\pi\chi_0)H_0$ and the magnetic field
strength $H_0$ practically coincide there.}; $\vector{b}(\vector{x},t)$ is
the magnetic field perturbation in the total field
 $\vector{B}(\vector{x},t)= \vector{B}_0 +\vector{b}(\vector{x},t)$;
$n_{e0}= \int d^3pf^{(0)}(\varepsilon_p)/(2\pi)^3$ is the density of the
electron gas. The pseudotensor $\Sigma_{kj}(\vector{x},t)= \mu_B\int
d^3pv_k\delta S_j(\vector{p}, \vector{x},t)/(2\pi)^3$ describes spatial
inhomogeneity of the magnetization and the last {\it new vector} term ($\sim c_A$)
corresponds to the parity violation for the evolution of the macroscopic
{\it axial} vector $m_j$ in SM.

Note that last two lines in the complicated spin RKE \pref{spin}
are the relativistic corrections and do not contribute to \pref{Bloch} in
NR plasma. Moreover, we omitted the small term
$- G_F\sqrt{2}c_A\chi_0[\vector{j}^{(\nu)}(\vector{x},t)\times
\vector{B}_0]_j$ originated from the third line (without
derivatives) in \pref{spin}  when comparing it with the third term in the
Bloch equation \pref{Bloch} stipulated by the standard spin precession in
QED plasma (arising due to the last term in the second line in
\pref{spin}).

We neglected also the fifth term in \pref{spin} that is
proportional to the vector coupling $\sim c_V$ since the background
polarization is small in our WKB approximation, $S^{(0e)}\sim \mid
(\mu_BB_0/\gamma)\mbox{\rm d}f^{(0e)}/\mbox{\rm d}\varepsilon_p\mid \ll
f^{(0e)}(\varepsilon_p)$, and this term is much less than the previous one
in \pref{spin} which is proportional to the axial coupling $\sim c_A$
retained in \pref{Bloch}.  This is in agreement with estimates of the
small polarization $<\lambda>= n_{e0}/n_e\sim 0.01$ made in
\cite{SemikozValle} (Nunokawa et al) even for strong magnetic field in SN
where $n_{0e}$ is the electron density on the main Landau level, $n_e$ is
the total electron density.

In order to break the
chain of hydrodynamical equations for magnetization moments, $m_j$,
$\Sigma_{ij}$, etc, one can consider long-wave spin waves with the wave
lengths $\lambda_s$ that are much bigger than the electron Larmour radius,
$\lambda_s\gg l_{eH}$.  In such case the spin perturbation function is
given by the first two moments:  $$ \mu_B\delta
S_j(\vector{p},\vector{x},t) = \frac{df^{(0)}}{d\varepsilon_p}\left[ \int
 \frac{d^3p}{(2\pi)^3} \frac{df^{(0)}}{d\varepsilon_p}\right ]^{-1} \times
\left (m_j(\vector{x},t) +
\frac{3v_i}{v^2}\Sigma_{ij}(\vector{x},t)\right)~, $$ that allows us to
complete the hydrodynamical system by the equation for the pseudotensor
$\Sigma_{ij}$, \begin{eqnarray} \label{sigma} \frac{\partial
\Sigma_{ij}(\vector{x},t)}{\partial t} + &&\Omega_e\Bigl (
e_{jtl}\hat{n}^B_l\Sigma_{it}(\vector{x},t)-
e_{itl}\hat{n}_l^B\Sigma_{tj}(\vector{x},t) \Bigr ) + \nonumber\\ && +
\frac{<v^2>}{3}\frac{\partial}{\partial x_i}\Bigl (m_j(\vector{x},t) -
\chi_0b_j(\vector{x},t)\Bigr ) - \nonumber\\&&-
\frac{G_Fc_A\mu_Bn_{e0}}{\sqrt{2}m_e}\int
\frac{d^3q}{(2\pi)^3}n_j(q)\frac{\partial \delta
f^{(\nu)}(\vector{q},\vector{x},t)}{\partial x_j} = 0~.
\end{eqnarray}
Here $\Omega_e = eB_0/m_e$ is the electron cyclotron frequency;\\
$<v^2> = [\int d^3pdf^{(0)}/d\varepsilon_p]^{-1}\int d^3p
v^2df^{(0)}/d\varepsilon_p$ is the average of the velocity squared;
$\hat{n}^B = \vector{B}_0/B_0$ is the unit pseudovector and last true
tensor term describes the parity violation through weak interactions. Note
that eq.  \pref{sigma} would be important accounting for the large spatial
dispersion $k<v>\sim \omega$ outside the region \pref{inequal} and claims
an inclusion of exchange interactions (from the ee-scattering exchange
Feynman diagram) we omitted here.

\section{Neutrino driven streaming instability of spin waves}
In this section we derive from generalized Bloch equation \pref{Bloch} the
dispersion equation in magnetized plasma in the presence of neutrino beam.

Apparently master RKE's for number density distribution functions
\pref{neutrino}, \pref{electron} should be consistent with the spin RKE
\pref{spin} and its consequence \pref{Bloch}. Therefore we are checking
below whether more general \pref{neutrino1}, \pref{electron2} are
necessary.  As we find below there is no difference between the use of
these equations for the spin wave propagation in plasma.

Let us consider for simplicity the case of long-wave perturbations
with the spectrum $\omega (\vector{k})$ obeying the inequalities
\begin{equation}
\label{inequal}
\frac{k^2}{\omega}\geq \omega\gg k<v>~.
\end{equation}
As the mean electron velocity $<v>$ is small in NR plasma, $<v>\ll 1$,
and keeping in mind the Maxwell equation written in the Fourier
representation $\vector{b} = [\vector{k}\times \delta \vector{E}]/\omega$
one finds from the condition $k/\omega\gg <v>$ that the electric field
contribution in the spin RKE \pref{spin}  ($\sim \delta \vector{E}$) is
negligible comparing with the magnetic one even for the maximum frequency
in whole space-like region $\omega\leq k$
relevant to the $\check{C}$erenkov resonance with neutrinos, $\omega =
\vector{k}\vector{n}$.  Without electric fields the transversal components
of the permeability and the susceptibility tensors ($i,j=x,y$) appearing
in the second term of \pref{Bloch} are diagonal in plasma,
$\mu_{ij}(\omega, \vector{k}) = \mu (\omega, \vector{k})\delta_{ij}$,
$\chi_{ij}(\omega, \vector{k}) = \chi (\omega, \vector{k}) \delta_{ij}$,
that differs this medium from ferromagnets.

Moreover, the latter inequality in
\pref{inequal}, $\omega\gg k<v>$, means the high-frequency approximation,
for which the spatial uniform susceptibility $\chi (t)$ and the
permeability $\mu (t) = 1 + 4\pi \chi (t)$ can be considered instead
of more general ones, $\chi (\vector{x},t)$ and $\mu (\vector{x},t )$, that
allows us to neglect the complicated pseudotensor term $\Sigma_{ij}$ in
\pref{Bloch} as well as the spatial dispersion in the
perturbations $m_{\pm}(\vector{x},t) = \int dt'\chi (t -
t')h_{\pm}(\vector{x}, t')$ and $b_{\pm}(\vector{x},t) = \int dt'\mu (t -
t')h_{\pm}(\vector{x},t')$ correspondingly where $h_{\pm}(\vector{x},t)=
h_x(\vector{x},t)\pm ih_y(\vector{x},t) $ is the magnetic field strength
perturbation.

Under such conditions we find from the linearized spin RKE \pref{spin}
the susceptibility in QED plasma neglecting neutrinos,
\begin{equation} \label{resonance} \chi (\omega) = \frac{\pm
\Omega_e\chi_0}{\omega \pm \Omega_e(1 - 4\pi \chi_0)}~, \end{equation} that
may be finite at the paramagnetic resonance $\omega= \mp
\Omega_e$ given by the equation $1 + 4\pi \chi (\omega)=0$. Latter follows
from the shortened Maxwell equation $\vector{k}\times (\vector{b} - 4\pi
\vector{m})=0$ when one neglects electric field terms.

Then substituting into
\pref{Bloch} the solution of the neutrino RKE \pref{neutrino} obtained in
the same linear approximation \pref{linear}, \begin{eqnarray}
\label{nulinear} &&\delta f^{(\nu)}(\vector{q}, \vector{k}, \omega) =
G_F\sqrt{2} c_V\frac{k_k}{(\omega -
\vector{k}\vector{n})}\frac{\partial f^{(\nu)}_0(q)}{\partial q_k}\int
\frac{d^3p_e}{(2\pi)^3}(1
- \vector{v}_e\vector{n})\delta f^{(e)}(\vector{p}_e, \vector{k}, \omega)
+\nonumber\\ &&+
\frac{G_Fc_A}{\sqrt{2}}\frac{k_kn_l(q) }{(\omega -
\vector{k}\vector{n})}\frac{\partial f^{(\nu)}_0(q)}{\partial q_k}\int
\frac{d^3p_e}{(2\pi)^3}\delta S_l(\vector{p}_e, \vector{k},
\omega)~,
\end{eqnarray}
we can rewrite the generalized Bloch equation \pref{Bloch}
in the Fourier representation as
\begin{eqnarray} \label{Fourier} - i\omega
m_j(\vector{k}, \omega) + && 2\mu_B[(\vector{m}(\vector{k},\omega) -
\chi_0\vector{b}(\vector{k}, \omega))\times \vector{B}_0]_j +
\frac{c_A^2\Delta_{\nu}}{2}A_l^{(\nu)}(\omega,
\vector{k})m_l(\vector{k},\omega)ik_j + \nonumber\\&& +
\frac{c_Ac_V\Delta_{\nu}k_j }{8\pi m_e}\left(A_i^{(\nu)}(\omega,
\vector{k}) - \frac{B^{(\nu)}(\omega,
\vector{k})}{\omega}k_i\right)\times\nonumber\\&&\times \Bigl
(e_{ikl}k_k[b_l(\vector{k},\omega ) - 4\pi m_l(\vector{k}, \omega)] +
\omega \delta \vector{E}_i \Bigr) = 0~.  \end{eqnarray} Here the factor
$\Delta^{\nu}$ is given after eq.  \pref{dispersion1}; the vector
$A_l^{(\nu)}\equiv A_l^{(\nu)}(\omega, \vector{k})$ and the scalar
$B^{(\nu)}\equiv B^{(\nu)}(\omega, \vector{k})$ depend on the neutrino
background distribution $\hat{f}_0^{(\nu)}(\vector{q})$,
\begin{equation}
\label{factorA} A_l^{(\nu)} = \int
\frac{d^3q}{(2\pi)^3}\frac{\hat{f}_0^{(\nu)}(\vector{q})}{q}\left(
\frac{[(\vector{k}\vector{n})^2 - k^2]n_l(q)}{(\omega -
\vector{k}\vector{n})^2} + \frac{(\vector{k}\vector{n})n_l -
k_l}{\omega - \vector{k}\vector{n}}\right)~,\end{equation}

\begin{equation}
\label{factorB} B^{(\nu)} = \int
\frac{d^3q}{(2\pi)^3}\frac{\hat{f}_0^{(\nu)}(\vector{q})}{q}\left(
\frac{[(\vector{k}\vector{n})^2 - k^2]}{(\omega -
\vector{k}\vector{n})^2}\right)~.\end{equation}
Obtaining last term in eq. \pref{Fourier} ($\sim c_Ac_V$) we used  the
exact Maxwell equation \pref{Maxwell2} when we substituted the {\it
convection current} $\delta j_i(\vector{k}, \omega) = \int d^3p v_i\delta
f^{(e)}(\vector{p}, \vector{k}, \omega)$ in the first line of eq.
\pref{nulinear}.

Note that the electromagnetic current $j_{\mu}^{(e)}(\vector{x},t)$
conserves automatically since we can neglect weak interactions in the
electron RKE \pref{electron} while retaining them in the neutrino RKE
\pref{neutrino}. In turn the neutrino influence the electron spin
evolution coming from the solution \pref{nulinear} is not changed if we
would substitute the solution of the complete equation \pref{neutrino1}
instead of the initial one \pref{neutrino}.

Really in the last case there appear the two additional terms in
\pref{nulinear},
\begin{equation}\label{additive}
- G_F\sqrt{2}c_V\frac{\partial f^{(\nu)}_0(q)}{\partial q_k}\delta
j^{(e)}_k(\omega, \vector{k}) - \frac{G_F}{\sqrt{2}}c_A\frac{\partial
f^{(\nu)}_0(q)}{\partial q_k}\delta m_k(\omega, \vector{k}) \end{equation}
which do not contribute to the generalized Bloch equation \pref{Bloch}.

One can
easily see that in NR plasma with a low electron density $n_{0e}\ll
m_e^3\sim 2\times 10^{31}~cm^{-3}$ the general condition of the
macroscopic description $2\pi/k\gg (n_0)^{-1/3}$ means that long wave
lengths exceeding the Compton one, $2\pi/k\gg m_e^{-1}$ are possible only,
and due to \pref{inequal} frequencies obey the inequality $\omega\leq k\ll
m_e$.

Accounting for low $\omega$ and $k$ and comparing
the term $\sim c_Vc_A$ with the previous one in eq.  \pref{Fourier}
($\sim c_A^2$) one finds that the convection current gives a negligible
contribution. Note that for
$\nu_{\mu,\tau}$-neutrinos with $\mid c_V\mid \ll 1$ such correction
becomes even less than for $\nu_e$.
Omitting also the small term $\sim \chi_0$ we arrive to the
shortened form of \pref{Fourier},
\begin{equation} \label{Fourier2} - i\omega
m_j(\vector{k}, \omega) + 2\mu_B[\vector{m}(\vector{k},\omega)\times \vector{B}_0]_j +
\frac{c_A^2\Delta_{\nu}}{2}A_l^{(\nu)}(\omega,
\vector{k})m_l(\vector{k},\omega)ik_j = 0~.  \end{equation}

Thus, from the generalized Bloch equation \pref{Fourier2}
we have derived finally the dispersion equation  in SM,
\begin{equation} \label{spectrum} (\omega^2 -
\Omega_e^2)\left(\omega - \frac{c_A^2\Delta^{(\nu)}}{2}A_z^{(\nu)}
k_z \right ) -
\frac{\omega^2c_A^2\Delta^{(\nu)}}{4}(A_-^{(\nu)}k_+ + A_+^{(\nu)} k_-) =
0~, \end{equation} with the vector $A_i^{(\nu)}$ as given in eq.
\pref{factorA}.

In the particular case of
the neutrino beam, $\hat{f}^{(\nu)}_0(\vector{q}) =
(2\pi)^3\delta^{(3)}(\vector{q} -\vector{q}_0)$, $\vector{n}_0 =
\vector{q}_0/q_0$, for the transversal neutrino propagation
$\vector{n}_0\perp \vector{B}_0$ with the beam direction along x-axis,
$\vector{n}_0= (1,0,0)$, and  the magnetic field $\vector{B}_0= (0,0,
B_0)$ we obtain from eq.  \pref{spectrum} the resonant excitation of spin
waves $\omega = \vector{k}\vector{n}_0 + i\delta = \Omega_e + i\delta$
with the increment $\delta$, \begin{equation} \label{increment} \delta
\simeq
\Omega_e^{1/3}\frac{\sqrt{3}}{4}\left(\frac{\Delta^{(\nu)}}{q_0}\right)^{1/3}
(\sqrt{2}\mid c_A\mid k\sin \theta_{q_0})^{2/3}\geq
\Omega_e\frac{\sqrt{3}}{4}\left(\frac{\Delta^{(\nu)}}{q_0}\right)^{1/3}
(\sqrt{2}\mid c_A\mid \sin \theta_{q_0})^{2/3} ~,
\end{equation}
where we substituted the scalar product $\vector{k}\vector{n}_0=k\cos \theta_{q_0}= k\sin \theta \cos \phi$
denoting $\theta_{q_0}$ as the angle between the neutrino beam direction
and the wave vector $\vector{k}$; $\theta$ is the angle of $\vector{k}$ with respect
to the magnetic field $\vector{B}_0$. Note that we relied in the last
inequality on the frequency approximation $k\geq \omega \simeq
\Omega_e$ assumed above.

Let us compare this increment with the fastest one in the case of
isotropic plasma following from the dispersion equation
\pref{dispersion1} \cite{Bingham},
\begin{equation}\label{isotropic}\delta_{weak} =
\frac{\sqrt{3}}{2}\omega_{pe}\left(\frac{\Delta_{\nu}}{q_0}\frac{\sin^6\theta_{q_0}}
{\cos^4\theta_{q_0}}\right)^{1/3}~.\end{equation} One can easily see the
advantage of the excitation of collective modes by the intense neutrino
flux in a polarized electron gas. For a strong magnetic field in a dense
plasma obeying $\Omega_e\gsim \omega_{pe}$ the angular dependence in eq.
\pref{increment} $\sim (\sin \theta_{q_0})^{2/3}$ gives a less suppression
of the increment for the small angles $\theta_{q_0}\leq \arccos (<v>/c)$
for which Landau damping for growing modes is absent. This is due to the
absence of the factor $(1 - \omega^2/k^2)^2$ in the new dispersion
equation \pref{spectrum}. It is obvious that such angular dependence
would be especially dangerous and important for the relativistic plasma
case.

There is the second advantage of spin waves enhanced via the weak axial vector
currents ($\sim c_{A}=\mp 0.5$) instead of \ the case of plasma waves
excited via the weak vector currents $\ $with the small vector coupling in the
case of muon and tau neutrinos (choosing the lower sign in $c_{V}=2\xi\pm0.5$
where $\xi\simeq 0.23$ is the Weinberg parameter ). This is the reason
why authors \cite{Bingham} considered the case of electron neutrinos only
and put for them $c_{V}\longrightarrow1$. Note that during the main
neutrino burst in SN all neutrino species are produced in the hot SN\ core
via the pair annihilation $e^{+}e^{-}\rightarrow\nu_{a}\tilde{\nu}_{a},$
$a=e,\mu,\tau$.

On the other hand, there are other arguments against streaming instability
driven by neutrino beams in SN (e.g. not collimated beam) \cite{Bento}
to be relevant for neutrino propagation in a magnetized medium.
Nevertheless, we have just showed that in a polarized electron gas the
dispersion equations are quite different from the case of the isotropic
plasma \cite{Bingham,Bento,Laming} that stimulates a future exploration of
collective plasma phenomena in the presence of intense neutrino fluxes.

\section{Discussion and conclusions}
The above derivation shows that the Bogolyubov method starting from QFT
Feynman diagrams remains a power and straightforward tool to obtain RKE's.
Being supported with an additional gauge restoring transformation similar
to \cite{Fujita} it leads to the same results obtained within Hamiltonian
approach of \cite{Bingham} for the case of neutrinos propagating in an
isotropic plasma.

In the case of a polarized electron gas the master RKE's derived above by
the Bogolyubov method allow to analyse a new phenomenon- spin wave
propagation in NR plasma enhanced by the neutrino beam. Note that complete
RKE's (\ref{neutrino1}), (\ref{electron2}) are not necessary to derive the
dispersion equation (\ref{spectrum}) for spin waves as we showed in
(\ref{additive}).

The violation of parity in the SM lepton plasma given by axial vector
currents ($\sim c_A$) leads to the growth of spin eigen modes through
their excitation by the intense neutrino flux. These spin waves are
generally coupled to the magnetosonic ones analogously to the case of
spin waves in ferromagnets \cite{SitenkoOraevsky} or can transfer their
energy to electromagnetic and plasma waves at the cross of spectra that
finally could lead to the heating of ions and the background plasma.

The possible explanation of the shock revival in SN by
different collective mechanisms including the neutrino driven streaming
instability seems to be very perspective goal for future studies in the
case of polarized electron gas.

The application of these mechanisms in the magnetized plasma behind the
shock and outside the SN neutrinosphere is self-consistent
with plasma parameters expected there.  Really,  we do not consider
neutrino collisions with matter within this region using the collisionless
neutrino RKE \pref{neutrino} and may also neglect the electron-ion
collision frequency $\nu_{ei}$
comparing with the cyclotron frequency at the paramagnetic resonance
$\omega = \Omega_e$.  Hence collisionless Vlasov approximation  should be valid
for the spin equation \pref{spin} as well as for the electron RKE
\pref{electron} since the Debye number is large, $N_D= n_{0e}r_D^3\gg
1$.  Indeed, in the field $B_0 =
10^{12}~Gauss$ the cyclotron frequency reaches $\Omega_e= 1.7\times
10^7B_0\sim 1.7\times 10^{19}~sec^{-1}$. This frequency is comparable with
the plasma one at the density $n_{e0}\sim 10^{29}~cm^{-3}$, $\omega_{pe}=
1.8\times 10^{19}~sec^{-1}$, and turns out to be larger than e.g. the
electron-proton collision frequency $\nu_{ep}=
50n_{e0}(cm^{-3})/(T_e(K))^{3/2}~sec^{-1}\sim 1.6\times 10^{17}~sec^{-1}$
in the surrounding NR plasma with the temperature $T_e\sim 10^9~K$. For
these parameters the Debye radius $r_D\sim 10^{-9}~cm$
corresponds to the large plasma parameter $N_D\sim 100\gg 1$.

Under such conditions for the mean neutrino energy
$q_0\sim 10~MeV$ the increment \pref{increment} reaches the maximum
value $\delta\sim 10^{10}~sec^{-1}$ as in previous optimistic
estimates\cite{Bingham, Bento}.  Note that this means too sharp collimated
neutrino beam with the spread of directions $\vector{n}_0$ for the fixed
wave number $k$ not exceeding very small value $\Delta
(\vector{k}\vector{n}_0)/\Omega_e\lsim \delta/\Omega_e\sim 10^{-9}$.  If
neutrinos move along radii beyond the neutrinosphere $r>R_{\nu}$ this
estimate is too optimistic since there is a spread of the angular
distribution of neutrino trajectories that damages simple model with
parallel rays assumed here \cite{Bento}.

Nevertheless, we think the simple dispersion equation \pref{spectrum}
based on the enhancement of pure magnetic field perturbations
is the only particular case of the general kinetic equations derived here.

Note that even in this approximation there are
some advantages of our model comparing with the isotropic
plasma case discussed in \cite{Bingham,Bento,Melrose,Laming} as we have
just shown in previous section (after eq. \pref{isotropic}).

A more general case with the overlap of electromagnetic eigen modes in  a
polarized medium and accounting for the spatial dispersion in such plasma
seems to be a more realistic model for a magnetized SN while it is beyond
of the scope of the present work .

We would like to acknowledge discussions with  A.S. Volokitin and A.I.
Rez.  This work was supported for V.S. by the RFBR grant 00-02-16271.


\end{document}